\documentclass[conference,onecolumn]{IEEEtran}
\let\labelindent\relax
\usepackage{cite}
\usepackage{graphicx}
\graphicspath{ {images/} }
\usepackage{wrapfig}
\usepackage{tcolorbox}
\usepackage{siunitx}
\usepackage{algorithm}
\usepackage{algorithmic}
\usepackage{enumitem}
\usepackage{multirow}
\usepackage{array}
\usepackage{mdframed}
\usepackage{subcaption}
\usepackage{stfloats}

\usepackage{url}

\newtcolorbox{sidebar}[1]{fonttitle=\bfseries,title=#1}

\newenvironment{halfsidebar}[1][r]
  {\wrapfigure{#1}{0.5\textwidth}\sidebar}
  {\endsidebar\endwrapfigure}

\begin{document}


\title{Using StackOverflow content to assist in code review}


\author
{
\IEEEauthorblockN{Balwinder Sodhi and Shipra Sharma}
\IEEEauthorblockA{Dept. of Computer Science and Engineering,
Indian Institute of Technology Ropar\\
Nangal Rd Ropar PB 140001 India\\
Email: \{sodhi, shipra.sharma\}@iitrpr.ac.in}
}
\maketitle


\begin{abstract}
An important goal for programmers is to minimize cost of identifying and correcting defects in source code. Code review is commonly used for identifying programming defects. However, manual code review has some shortcomings: a) it is time consuming, b) outcomes are subjective and depend on the skills of reviewers. An automated approach for assisting in code reviews is thus highly desirable.

We present a tool for \textit{assisting in} code review and results from our experiments evaluating the tool in different scenarios. The tool leveraged content available from professional programmer support forums (e.g. StackOverflow.com) to determine potential defectiveness of a given piece of source code. The defectiveness is expressed on the scale of \{\texttt{Likely defective, Neutral, Unlikely to be defective}\}.

Basic idea employed in the tool is to: \textit{a)} Identify a set $P$ of discussion posts on StackOverflow such that each $p\in P$ contains source code fragment(s) which sufficiently resemble the input code $C$ being reviewed. \textit{b)} Determine the likelihood of $C$ being defective by considering all $p\in P$. A novel aspect of our approach is to use \textit{document fingerprinting} for comparing two pieces of source code. Our choice of \textit{document fingerprinting} technique is inspired by source code plagiarism detection tools where it has proven to be very successful. 
In the experiments that we performed to verify effectiveness of our approach source code samples from more than 300 GitHub open source repositories were taken as input. A precision of more than 90\% in identifying correct/relevant results has been achieved.

\end{abstract}

\begin {IEEEkeywords}
Code Review, StackOverflow, Software Development, Crowd Knowledge, Automated Software Engineering
\end{IEEEkeywords}

\section{Introduction}
We present a novel tool that assists in carrying out effective code reviews. Identifying and fixing buggy code consumes significant time and resources in a software development project. Code reviews by peers\cite{sommerville2010software} and experienced programmers is an effective method \cite{mcconnell1993code,huizinga2007automated} for identifying potentially buggy code. However, the process of code review is slow and quality of results depends on skills and experience of the reviewers involved. Moreover, a code review carried out by an individual expert is always subjective and hence open for questioning. An automated tool which can improve the quality of code reviews is thus highly desirable. Especially if such a tool can reduce or eliminate the subjectivity associated with individual experts' code reviews, it will be a considerable gain. This is exactly what our tool helps with by leveraging ``crowd expertise''.

\textit{How do programmers acquire ``expertise'' to become ``expert reviewers''?}\\
Mostly one learns from other's experience which may be available in variety of forms such as in a text book, a best practices guide or available in the on-line question and answer forums. It is observed that often times a programmer, when faced with a problem or a bug in some source code, turns to searching an on-line professional programmer forum such as StackOverflow\textsuperscript{\textregistered} for assistance. In fact, most software vendors now use StackOverflow\footnote{Interest in StackOverflow.com has been steadily increasing over the last decade: https://trends.google.com/trends/explore?date=all\&q=\%2Fm\%2F05mw61p} as a programmer support channel for their software. As such the information available on professional programmer forums has become a valuable source of experiential knowledge -- or ``crowd expertise'' -- about several aspects of software design and development. 

\textit{Challenges in using ``crowd expertise'' for assisting in software development tasks}\\
Search engines and information retrieval (IR) technologies have made it easy to locate relevant information on professional programmer support forums, however, in order to zero-in on a suitable solution for the problem at hand a programmer has to manually sift through the content presented by IR tools. One of the reasons why a programmer needs manual sifting through the search results is that normal IR tools may not always take into account the semantic context in which a programmer is operating. As such the ability to derive benefit from the content available on professional programmer support forums is limited by the domain expertise of a programmer and his/her fluency in relevant technical vocabulary.

To address the issues concerning consumption of raw ``crowd expertise'' researchers have leveraged\cite{nlp-survey-2014,nlp-application-2016} technologies such as Natural Language Processing (NLP) and Knowledge Discovery (KD). Particularly, the application of NLP techniques for deriving sentiment (on a selected scale) has been a popular direction. However, even the existing text analysis techniques such as CoreNLP\cite{corenlp2014, socher2013recursive}, Vader\cite{hutto2014vader} etc. may not give accurate results when used for determining defectiveness ``sentiment'' about a piece of source code by analysing the narrative associated with the source code. This is because such text analysis techniques, with their commonly used models, seem to perform poorly when applied to domain specific narrative. For example consider the example post content shown in the Sidebar-1.
\begin{halfsidebar}{Sidebar-1 (Example post)}
\begin{quote}
	\textit{Consider the following code. This function always goes into infinite loop at sixth line even for the normal inputs.}\\
    \begin{verbatim}
1.  check = True
2.  count = 1
3. 
4.  def flagged_sum(flag, count):
5.    sum = 0
6.    while check or flag:
7.        sum = sum + count
8.        print "OK\n"
9.    return sum
10.
11. flagged_sum(True, 2)
    \end{verbatim}
\end{quote}
\end{halfsidebar}

It is highly unlikely that a comment such as the narrative in above example about a source code will be  considered positive. However, if you run the above narrative through a sentiment analysis tools such as \cite{corenlp2014, hutto2014vader} they report the narrative sentiment as ``positive'', which is misleading if defectiveness of the code that accompany the narrative is judged from the sentiment of that narrative.

In this paper we present a system which bridges such gaps. To understand general working approach of our system consider the following (very simplified) scenario involving professional programmer support forum such as StackOverflow. Suppose that a programmer has encountered a subtle bug in his/her program. He/she then seeks assistance on StackOverflow by describing the issue in a post, $p$, there. In the post he/she has attached the relevant source code, $c$, from his/her program. Suppose this question has also been up-voted sufficient number of times by viewers, thus confirming the validity of issue described in the question. Now, if another piece of source code, $c'$, which is being reviewed, sufficiently resembles $c$, then by association one can infer that $c'$ is \textit{highly likely} to encounter the same issue as reported in the post $p$.

The proposed tool works by identifying discussion posts on StackOverflow such that each post, $p$, contains source code, $c_p$, which sufficiently resembles the input code, $c_{input}$, being reviewed. By analyzing the content (text narrative, meta-data and attached source code $c_p$) of the post $p$ we determine whether $c_p$ represents a defective code or not. Because $c_p$ sufficiently resembles $c_{input}$, we can infer the defectiveness of $c_{input}$ itself. Table-\ref{tab:defectiveness_scale} depicts the scale used for specifying/measuring code's defectiveness scores.

\begin{table}
	\centering
	\caption{Scale for defectiveness score, $\alpha$}
	\label{tab:defectiveness_scale}
	\begin{tabular}{r|l}
		\hline
		\textbf{Value of $\alpha$} & \textbf{Meaning} \\ \hline
		-1 & Likely to be defective \\
		0 & Neutral \\
		1 & Not likely to be defective \\
		\hline
	\end{tabular}
\end{table}

The proposed system thus \textit{assists in} code reviews and identification of potentially defective source code. It improves the confidence in a code review by identifying similar scenarios on professional programmer support forums such as StackOverflow.

\textbf{Paper is organized as follows:} We discuss the related work in Section-\ref{sec:related_work}. Design of the proposed tool is presented in Section-\ref{sec:proposed_approach} and its implementation details in Section-\ref{sec:impl}. In Section-\ref{sec:evaluation} we provide in-depth discussion on empirical evaluation of our tool and present key observations in Section-\ref{sec:key_obs}. We have also discussed in Section-\ref{sec:threats} some threats to the validity of our approach.

\subsection{Related work}
\label{sec:related_work}
The use of  \textit{crowd expertise} for addressing software engineering issues is not entirely new. There have been recent works such as \cite{Ponzanelli2014, Seahawk2013} which aim to assist programmers by leveraging content available on professional programmer support forums. Both of these are recommender systems which analyze the source code being typed in an IDE, and generate suitable queries on-the-fly to retrieve relevant discussion threads from StackOverflow. \cite{Ponzanelli2014} proposes a ranking model to select the most relevant discussion posts from StackOverflow based on the current context (i.e. the source code being typed) in the IDE. \cite{Seahawk2013} is a more primitive version of \cite{Ponzanelli2014}. They both use Apache Solr\footnote{http://lucene.apache.org/solr/} document indexing engine to index the StackOverflow posts that are retrieved from the dump of StackOverflow.

Another related approach has been proposed by \cite{Thummalapenta:2007:PPA:1321631.1321663} which suggests API invocation sequences to programmers who are looking to obtain a target object starting with a different object. They depend on code search engines to retrieve relevant code samples from the Web, and then statically analyze such code snippets to identify the method invocation sequences. Yet another novel approach that assists a programmer by suggesting him/her the code snippets is presented in \cite{mishne2012typestate}. It suggests to a programmer how an API should be used in a given programming context.

Similarly, \cite{rahman-roy2014} proposes a \textit{``Context-Aware IDE-Based Meta Search Engine for Recommendation about Programming Errors and Exceptions''}. They implemented an IDE (Eclipse) plug-in that makes use of the APIs provided by popular web search engines (viz. Google, Yahoo, Bing) to search the entire Web using keywords from a programming context in an IDE. They  determine relevant results by taking into consideration the programming context as well as popularity of the candidate search results.

There have also been works such as \cite{Cubranic2003} which infers possible links/relatedness among artifacts of a software development project by analyzing the projects artifact repositories.

Main goal in case of most the existing works such as the ones highlighted above is to suggest code samples or relevant Q\&A discussions which can assist the programmer in completing a coding task in an IDE. Emphasis is on assisting a programmer in \emph{writing the code} by leveraging code snippets (or discussions about code) from different sources. Secondly, most of the existing works leverage standard information retrieval (IR) techniques to fetch relevant code/content from various sources such as the Web. For instance, \cite{Ponzanelli2014, Seahawk2013} make use of \textit{term frequency--inverse document frequency} based text mining techniques, whereas \cite{Thummalapenta:2007:PPA:1321631.1321663, rahman-roy2014} make use of existing search engines/APIs for retrieving relevant content.

Specifically, in the area of code review there have been sentiment analysis of reviews \cite{Ponzanelli2014}, study on parameters that affect code review \cite{Seahawk2013}, and semi-automation in the review process \cite{di2014ontology}. Here, although \cite{Ponzanelli2014} differs from our work as it just reviews the comments of three OSS and characterizes them on the basis of their sentiment and \cite{di2014ontology} proposes a tool based on static analysis of code to automate the code review process. \cite{mcgraw2008automated} gives a detailed description of various (semi-) automated tools of code reviews avialble for use but concerning specifically to security aspect of the software. These tools are based on static analysis and their performance depends on code quality. 

Main aim of the work presented in this paper, however, is to assist in reviewing a given piece of source code. The proposed tool supports the reviewer by identifying relevant content on professional programmer support forums. Secondly, instead of using regular IR approach to identify matching code present in Web content, we leverage a proven and robust \textit{document fingerprinting} technique called Winnowing\cite{schleimer2003winnowing} to identify the matching code. Winnowing has been used in one of the most successful code plagiarism detection tools MOSS\footnote{\url{https://theory.stanford.edu/~aiken/moss/}}. 

\begin{algorithm}
	\caption{Steps in our approach\label{alg:main_algo}}
	\begin{algorithmic}[1]
		\REQUIRE $PostsDB :=$ Database containing posts data (i.e. code, meta-data, text, fingerprints etc.).
		\ENSURE $\alpha :=$ Overall defectiveness score for an input source code unit/file.\\
		$P :=$ Ranked set of relevant posts supporting the obtained probability value. \\
		
		\STATE $C :=$ Set of self-contained code blocks (strings) in a source code unit.
		\STATE $\Delta l :=$ Code size error tolerance to use when searching for fingerprint matches in $PostsDB$. It is \% of input code string length.
		\STATE $\delta :=$ Fingerprint match threshold to use when searching in $PostsDB$. It is \% of fingerprint length of the input code.
		\STATE $\alpha :=$ 0 /*\texttt{ See Table-\ref{tab:defectiveness_scale} }*/
		\FOR {Every code block string $c$ in $C$}
		\STATE $F_c :=$ Fingerprint calculated on $c$\\ 
		~~~~~~~/* \texttt{$F_c$ is a set of hash values.} */
		\STATE $l_c :=$ Length of code string $c$.
		
		\FOR {Every record $r$ in $PostsDB$}
		\IF{($\left\vert{l_{r.code} - l_c }\right\vert \leq \Delta l$) AND ($\left\vert{r.fingerprint \cap F_c}\right\vert \geq \delta$)}
		\STATE $P := P \cup \lbrace r \rbrace$
		\STATE $AggregateDefectivenessScore(\alpha, r.defScore)$
		\ENDIF
		\ENDFOR
		\ENDFOR
		\STATE Sort $P$ to select top $k$ matches having highest degree of code match.
	\end{algorithmic}
\end{algorithm}

\section{Proposed approach}
\label{sec:proposed_approach}
Association between software development and crowdsourced knowledge has been studied and confirmed by Bogdan et. al.\cite{stackoverflow-github} where they studied data from GitHub (a popular repository of open source software) and StackOverflow. The type of questions that are asked and get answered or remain unanswered on StackOverflow has been explored by \cite{stackoverflow-topics, stackoverflow-empirical-study}. One can identify primarily three types of posts on StackOverflow:
\begin{enumerate}[label=Type-\arabic*, leftmargin=*]
	\item Questions posted by programmers soliciting help/solution for a programming problem that they are facing with code/API etc.
	\item Questions posted by programmers soliciting recommendations/suggestions about a design decisions choice. For example, whether one should use MongoDB or MySQL as data store in a specific application development scenario.
	\item Responses posted by programmers to other's questions of the above types.
\end{enumerate}

We also find that there are more than 1.5 times\footnote{On StackOverflow, there are more than 14m questions and 22m answers as of July 2017. See https://data.stackexchange.com/.} as many posts of Type-3 as there are of Type-1 and 2 combined. Further, we find\footnote{Our query for finding this number is available here: https://data.stackexchange.com/stackoverflow/query/557623.} that more than 73\% of Type-1 posts contains source code snippet(s). Each of such posts typically describes some problem involving the source code snippet included in the post. 

In view of the above it can be argued that: i) A piece of code accompanying a StackOverflow question is quite likely to be involved in a defect\cite{stackoverflow-topics, stackoverflow-empirical-study}, and ii) The code accompanying accepted or high scoring answers to such a question is quite likely to be free from the problem described in associated question post.

Our approach relies on the above results. Main idea of our approach can be stated as follows:
\begin{itemize}
	\item Professional programmer support forums such as StackOverflow contain numerous posts each containing the description of a software program related problem. Each such post, $p_{question}$, typically contains relevant source code, $c_{defective}$, in full or in part.
	\item Associated with each post $p_{question}$ there are zero or more reply posts, $p_{reply}$, which potentially solve the problem described in $p_{question}$. The post $p_{reply}$ also may contain source code, $c_{correct}$, which is a correct version of $c_{defective}$ or somehow solves the problem(s) described in $p_{question}$.
	\item If a piece of source code, $c_{reviewed}$, is to be reviewed and it resembles sufficiently with the code present in a post $p_{question}$ or $p_{reply}$ then by association we can infer about the quality of $c_{reviewed}$. Depending on whether $c_{reviewed}$ matches significantly with $c_{defective}$ or with $c_{correct}$, we may deduce whether $c_{reviewed}$ is likely to be defective or not.
\end{itemize}

\begin{sidebar}{Sidebar-2 (Winnowing Algorithm for Document Fingerprinting)}
Basic goal of the Winnowing algorithm for document fingerprinting is to find substring matches between a given set of document such that they  satisfy two properties:
\begin{enumerate}
\item All substring matches that are at least as long as some threshold, $t$, are guaranteed to be detected.
\item Any matches shorter than the noise threshold, $k$, are not detected.
\end{enumerate} 
The constants $t$ and $k \leq t$ are chosen by the user. High-level steps in the algorithm are as follows:
\begin{enumerate}
\item Take the document text $T$ to be fingerprinted. E.g. \texttt{A do run run run, a do run run.}
\item Transform the $T$ to $T'$ by removing irrelevant features. E.g. \texttt{adorunrunrunadorunrun}
\item Generate a sequence of $k-$grams from $T'$. E.g. A sequence of 5-grams: \texttt{adoru dorun orunr runru unrun nrunr runru 
unrun nruna runad unado nador adoru dorun orunr runru unrun}
\item Generate hashes for each of the $k$-grams. E.g. \texttt{77 74 42 17 98 50 17 98 8 88 67 39 77 74 42
17 98} 
\item Identify windows of hashes of length $m$.
E.g. Windows of length 4 for the above example text are: \texttt{(77, 74, 42, \underline{17}) (74, 42, 17, 98)
(42, 17, 98, 50) (17, 98, 50, \underline{17})
(98, 50, 17, 98) (50, 17, 98, \underline{8})
(17, 98, 8, 88) (98, 8, 88, 67)
( 8, 88, 67, 39) (88, 67, \underline{39}, 77)
(67, 39, 77, 74) (39, 77, 74, 42)
(77, 74, 42, \underline{17}) (74, 42, 17, 98)}
\item In each window select the minimum hash value. If there is more than one hash with the minimum
value, select the rightmost occurrence. Now save all selected hashes as the fingerprints of the document. E.g. \texttt{\{17 17 8 39 17\}}
\end{enumerate}

\end{sidebar}

\subsection{Design}
\label{sec:design}
Our approach is aimed at reviewing source code files, one at a time, that a programmer creates using an IDE or a text editor. Code blocks from source code to be reviewed are used as inputs to a \textit{document fingerprinting}\cite{schleimer2003winnowing} driven search algorithm as described in Algorithm-\ref{alg:main_algo} which does the following:
\begin{itemize}
	\item Identifies posts that contain source code resembling \textit{sufficiently} with the input.
	\item Determine the defect probability for the code attached with each such post. We use the steps as described in Algorithm-\ref{alg:defectiveness} for this task.
	\item Use the above probability scores to assess the defect probability for input code.
\end{itemize}

A key idea employed in our approach is to calculate ``fingerprints'' for the code which is attached with posts. Later we search for relevant posts using these fingerprints. Fingerprinting of code is done using Winnowing\cite{schleimer2003winnowing} which is a local algorithm for ``document'' fingerprinting. Overall process is described in Algorithm-\ref{alg:main_algo}, and the logical structure of the overall system is shown in Fig. \ref{fig:arch}.

\subsection{Deriving defectiveness score for source code attached with a post}
\label{sec:deriving_defectiveness}
Each post on StackOverflow contains useful meta-data (discussed further in Section-\ref{sec:datadump}) in addition to text narrative and attached source code. Interesting meta-data attributes of the post are:
\begin{enumerate}
	\item \texttt{Score}: It is an integer value that reflect how many viewers have endorsed or disapprove the post. Each endorsement increments the score while a disapproval decrements it. Higher value of score for a post indicates the genuineness of its contents.
	\item \texttt{ViewCount}: It is an integer value which reflects how many times this post has been viewed by users. ViewCount of a post can be taken to represent prevalence of the issues being discussed in the post.
\end{enumerate}

To derive the defectiveness score $\alpha$ for a post we consider: a) sentiment of the post's narrative text as computed with NLP tools, b) the \texttt{Score} and \texttt{ViewCount} values for the post . Algorithm-\ref{alg:defectiveness} shows the steps involved in determining $\alpha$ for the code present in a given post. Function \texttt{CalculateNarrativeSentiment()} makes use of CoreNLP\cite{corenlp2014} to determine the sentiment score $s_{narrative}$ for text of the post. Function \texttt{CalcDefectivenessScore()} applies $s_{narrative}$ to $\alpha$ in order to arrive at final value of the defectiveness score of the post.

\underline{Choice of $s_{threshold}$} is made by considering statistical spread of \texttt{Score} values for Q\&A posts available on StackOverflow. We observed\footnote{Our query is available here: \url{https://data.stackexchange.com/stackoverflow/query/759376}} that the average value of \texttt{Score} for questions is 1 and that for replies is 2. The standard deviation for both being about 20 indicates that the \texttt{Score} values are fairly spread out. Therefore, to be on the conservative side, we chose the value of $s_{threshold}$ to be 1.

\begin{algorithm}
	\caption{Calculating defectiveness score $\alpha$ for the code present in a post.	\label{alg:defectiveness}}
	\begin{algorithmic}[1]
		\REQUIRE
		$M :=$ Set of meta-data attributes for a post.\\
		$T :=$ Text narrative present in the post.\\
		$s_{threshold} :=$ Threshold for minimum score for a post.
		\ENSURE
		$\alpha_p :=$ Defectiveness score for code present in a post, $p$.\\
		
		\STATE $\alpha_p \leftarrow 0$ /*\texttt{ 0 means neutral. See Table-\ref{tab:defectiveness_scale} }*/
		\IF {$M.PostType = QUESTION $ AND $M.Score > s_{threshold}$}
		\STATE $\alpha_p \leftarrow -1$
		\ELSIF {$M.PostType = ANSWER$} 
		\IF{$M.Score > s_{threshold}$}
		\STATE $\alpha_p \leftarrow +1$
		\ELSE
		\STATE $\alpha_p \leftarrow -1$
		\ENDIF
		\ENDIF
		\STATE $s_{narrative} \leftarrow CalculateNarrativeSentiment(T)$
		\STATE $\alpha_p \leftarrow CalcDefectivenessScore(\alpha_p, s_{narrative})$
		\RETURN $\alpha_p$
	\end{algorithmic}
\end{algorithm}

\section{Implementation}
\label{sec:impl}
Here we describe the salient features of our implementation of the approach presented in previous section (i.e. Section-\ref{sec:proposed_approach}). The necessary software program for implementation of our proposed design has been developed using Java programming language. Java was chosen primarily because of: \textbf{a)} availability of expertise in Java with us. \textbf{b)} Ease of integrating with variety of database engines and other middleware. \textbf{c)} Reasonably good performance for the effort devoted to coding.

\subsection{Details of processing flow}
\label{sec:processing_flow}
The proposed tool relies on availability of certain data artifacts for its correct functioning. These data artifacts are derived from the StackOverflow data dump. Once the data artifacts are setup, the tool can serve code review requests. Thus, broadly, there are two tasks in the processing flow:
\subsubsection{Data preparation}
Main task here is to extract the relevant information from posts available in downloaded data dump of StackOverflow. The extracted information includes post's meta-data, textual description of the problem as present in a post and source code snippets if present in the post's body. In addition to meta-data we also need to calculate fingerprints and defectiveness score for the source code snippets extracted from a post. We also extract the information about links (i.e. related and duplicate posts) for a given post. Links/relatedness data is available in StackOverflow data dump itself. Finally all this information is persisted in a database.

\subsubsection{Code review requests handling}
Following are the main steps:

\begin{enumerate}[label=\roman*)]
	\item Take the source code input by a user. A user may supply a single file or a directory containing multiple files. Each file is processed independently.
	\item For each input source file perform the steps as described in Algorithm-\ref{alg:main_algo}.
\end{enumerate}

\subsection{Main constituent elements}
\label{sec:constituent_elements}

\begin{figure*}
	\centering
	\begin{subfigure}[ht]{0.5\textwidth}
		\centering
		\includegraphics[width=0.9\linewidth]{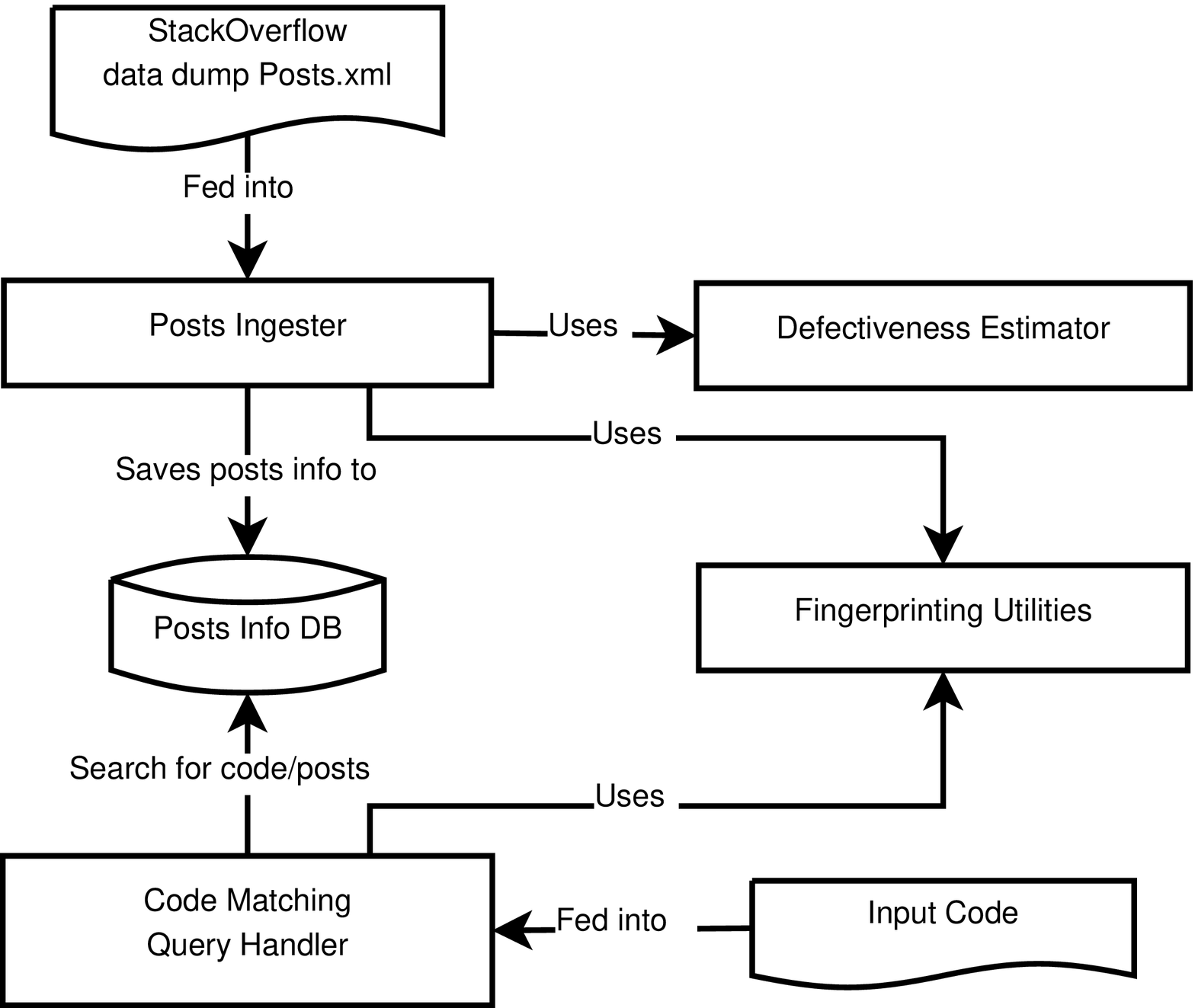}
		\caption{Logical structure of the system.}
		\label{fig:arch}
	\end{subfigure}%
	\begin{subfigure}[ht]{0.5\textwidth}
		\centering
		\includegraphics[width=0.7\linewidth]{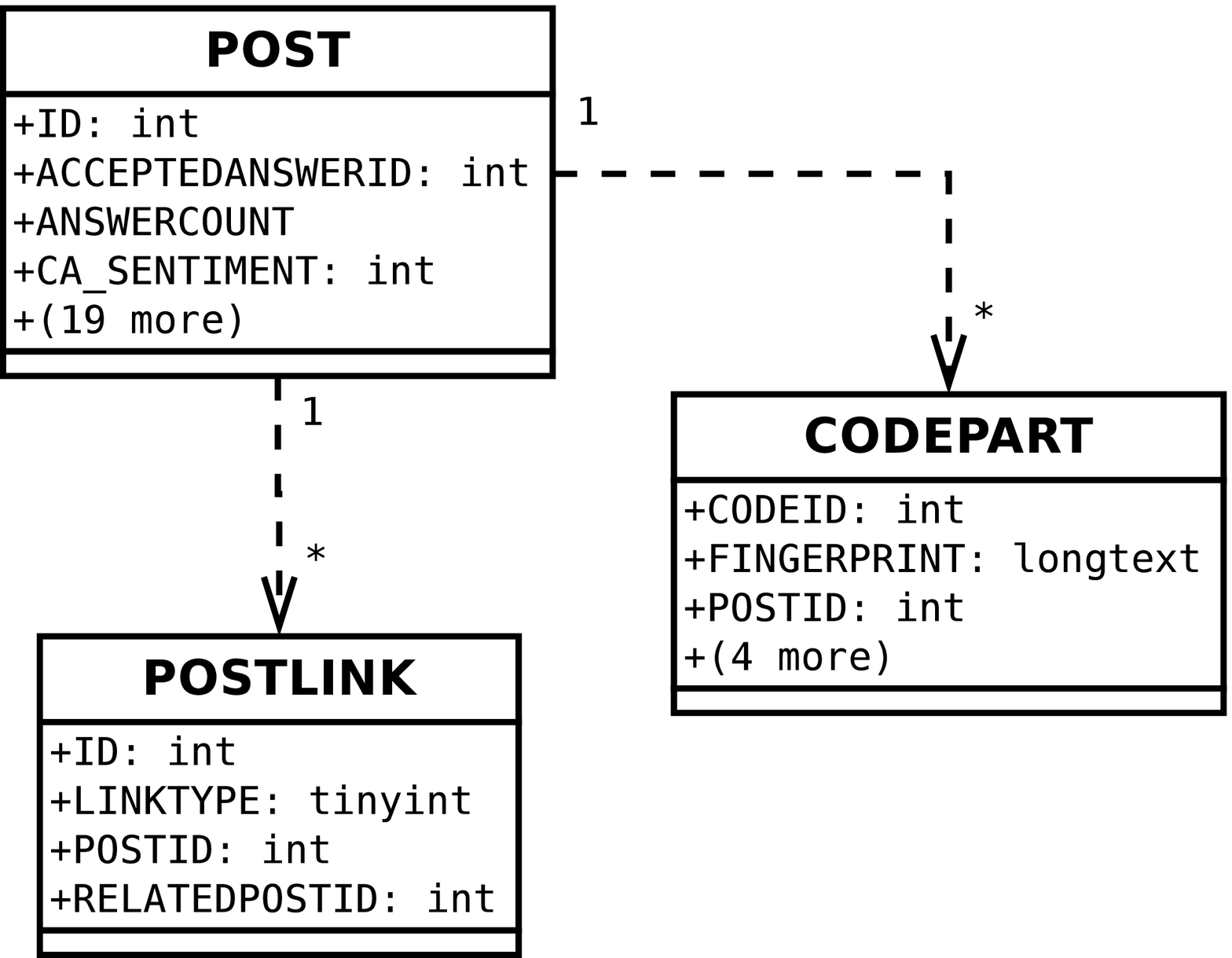}
		\caption{Schema of PostsDB.}
		\label{fig:posts_db}
	\end{subfigure}
	\caption{Structure of the complete system and Schema of our database}
	\label{fig:sys_arch}
\end{figure*}

Overall architecture of the system is shown in Fig. \ref{fig:arch}. Following elements are the main constituents of our implementation:
\subsubsection{StackOverflow Data Dump}
\label{sec:datadump}
Content contributed by professional programmers on Stackoverflow.com is made available to general public under Creative Commons License. We downloaded a snapshot of posts from \texttt{\url{https://archive.org/download/stackexchange/stackoverflow.com-Posts.7z}} (file is $\approx 9.2 GB$ in size at the time of download). This archive contains an XML file named Posts.xml whose structure looks like:

\begin{verbatim}
<posts>
<row ... />
<row ... />
..
</posts>
\end{verbatim}

Detailed description of each XML element and its attributes is available here: \texttt{\url{https://ia800500.us.archive.org/22/items/stackexchange/readme.txt}}.
The most relevant attributes of the \texttt{row} element for us are: \texttt{Id, PostTypeId, Score, ViewCount, Body, Title} and \texttt{Tags}. Values of \texttt{Score} and \texttt{ViewCount} are useful for deriving the sentiment value of the post. Code fragments to be fingerprinted are present inside the value of \texttt{Body} attribute. Value of \texttt{Tags} is useful for filtering the query results when searching for posts with specific tags.

\subsubsection{Posts info database (PostsDB)}
\label{sec:postsdb}
Information extracted from StackOverflow posts is persisted in an RDBMS for querying by the tool. Schema of such a database is shown in Fig. \ref{fig:posts_db}. It consists of three tables:
\begin{enumerate}[label=\roman*)]
	\item \texttt{POST}: This table stores the meta data found in a post.
	\item \texttt{CODEPART}: This table stores the calculated attributes such as the code and text parts extracted from a post, code size, fingerprint for the code etc.
	\item \texttt{POSTLINK}: Information about any links among different posts is held in this table. For instance, the information whether a post is duplicate of another post is kept here.
\end{enumerate}

\subsubsection{Posts Ingester}
\label{sec:ingester}
Purpose of this module is to extract relevant information from posts found in StackOverflow data dump and persist in the \texttt{PostsDB}. Relevant information includes entities such as: \textit{a)} embedded source code fragments in the posts, \textit{b)} fingerprints of the embedded source code and defectiveness score for the post and \textit{c)} meta data such as \texttt{ViewCount, Score} etc. available with each post. 

The information which this module extracts and persists in \texttt{PostsDB} will be searched by the \texttt{Code Matching Query Handler} (see Sec. \ref{sec:finder}) module when looking for matches with input source code being reviewed.

\paragraph*{\underline{Determining the input code size threshold}}
StackOverflow posts also contain very short pieces of text that are marked as code. Examples are function or variable names etc. Such small pieces of code are not useful for our tool. In order for the \texttt{Code Matching Query Handler} to produce meaningful results only those posts which have source code above certain minimum size are useful. Therefore the \texttt{Post Ingester} persists data of only those posts that contain source code whose size is above certain minimum value. 

The \texttt{Code Matching Query Handler} module uses a self-contained code block (chosen as a function definition in most programming languages) as the input for its code search operation. Therefore, the size of such a self-contained code block is used as a threshold when ingesting posts data into \texttt{PostsDB}. For determining value of this threshold we considered two aspects concerning how programmers typically write source code:
\begin{enumerate}[label=\roman*)]
	\item The coding best practice recommendations\footnote{A popular guide from Google Inc. is: \url{https://google.github.io/styleguide/}} across different programming languages, and
	\item Empirically observed size of code fragments present in the StackOverflow posts.
\end{enumerate}

A common best practice recommendation for writing functions/methods in most programming languages states that length of a function body be such that a programmer can view the entire function within a single page/screen, and that each line of code does not spill beyond 80 columns/characters on screen. Considering the common display resolutions, size of such a function works out to be about 40 lines (each 80 characters long) of code. Assuming that on average 20\% of a line is whitespace, the total non-whitespace characters length for a function works out to be $40 \times 80 \times (1-0.20) = 2560$ characters.

Table \ref{tab:posts_stats} shows distribution of code fragment sizes that were found in the \texttt{PostsDB}. Observed average size (2216) of a code fragment is in close vicinity of the value (2560) obtained based on best practices parameters. However, as shown in Table \ref{tab:posts_stats}, the standard deviation of code size is very close to the mean code size. This implies that the code sizes have large variation and are not centered around the mean. Therefore, the input code size for running evaluation experiments can be safely chosen to be within any range that gives adequate number of samples. Input code size in our experiments lies in 1 -- 10kB range.

\begin{table}
	\centering
	\caption{Code size statistics for our posts sample.}
	\label{tab:posts_stats}
	\begin{tabular}{l|r}
		\hline
		ITEM & VALUE \\ \hline
		Number of posts considered, $N$ & 8339939.00 \\
		Average code size (char), $c_{avg}$ & 2216.01 \\
		Max. code size (char), $c_{min}$ & 41802.00 \\
		Min. code size (char), $c_{max}$ & 1001.00 \\
		Std. dev. of code size (char), $c_{sd}$& 2110.38 \\
		\hline
	\end{tabular}
\end{table}

\subsubsection{Code Matching Query Handler}
\label{sec:finder}
Main purpose of this module is to identify relevant posts based on the input source code and the source code fragments found in the posts. Algorithm-\ref{alg:main_algo} outlines the processing steps performed by this module. 

It can take input in the form of a directory containing multiple source code files or as a single source code file. By default this module breaks an input source file into a collection of self-contained code blocks. Such a code block here typically is a function/method definition. It uses one code block at a time to search for matches in the \texttt{PostsDB}. Overall result for a given input source file is determined by combining the results for each self-contained code block. 

One problem that this module has to handle is that the code which we have extracted from StackOverflow posts is not always in the form of self-contained code blocks such as a complete function/method definition. In many cases the code that we find in a post consists of one or more snippets taken out of function definition(s). Luckily, the document fingerprinting method (see Winnowing Algorithm side box in Section-\ref{sec:design}) that we use for comparing the source code does not impose any structural/formatting requirements on the pieces of code being compared. Therefore our tool automatically takes care of the problem.

\subsubsection{Fingerprint Utilities}
\label{sec:fputils}
In order to search the relevant posts, source code matching is performed using \textit{document fingerprints}. Main purpose of this module is to provide implementation of document fingerprinting functions using Winnowing\cite{schleimer2003winnowing} algorithm. An illustration of the Winnowing algorithm is presented in the side box in Section-\ref{sec:design}.

\subsubsection{Defectiveness Estimator}
\label{sec:sentiment_analyzer}
This module implements our approach for calculating the defectiveness score, $\alpha$, for a piece of source code which is attached with a StackOverflow post. In order to estimate the value of $\alpha$ for a StackOverflow post $p$ we make use of $p$'s meta-data and the narrative text description from \texttt{PostsDB}. The narrative text description of $p$ is used for deriving natural language based sentiment about the post via CoreNLP\cite{corenlp2014}. Table-\ref{tab:defectiveness_scale} shows the scale used for $\alpha$.

Section-\ref{sec:deriving_defectiveness} outlines the approach for estimating value of $\alpha$, and Algorithm-\ref{alg:defectiveness} describes the calculation steps.

\section{Evaluating effectiveness of the tool}
\label{sec:evaluation}

The proposed tool can be considered effective if it forms its code review assessment using sufficiently relevant posts, and limits the use of irrelevant posts in its assessment. Once the correct posts which match the input source code have been identified by the tool, estimating defectiveness score of the posts and input code becomes relatively easier (Algorithm-\ref{alg:defectiveness}). Therefore, one of the goals of our evaluation experiments has been to evaluate the performance of our tool in identifying correct posts. In order to determine the efficacy of our tool we used multiple approaches. 

First we assessed tool's efficacy based on the quality of code matching  (Section-\ref{sec:code_matching_accu}). Secondly, we evaluate the proposed tool by using crowd-sourced pairs of $\langle original, duplicate\rangle$ posts (Section-\ref{sec:verif_dup_posts}). Finally, the tool was also evaluated (Section-\ref{sec:verif_manual}) by a group of experienced programmers on real projects.

\begin{algorithm}
	\caption{Setting up verification data.\label{alg:verification_setup}}
	\begin{algorithmic}[1]
		\REQUIRE
		PostsDB $:=$ Database containing posts data (i.e. code, meta-data, text, fingerprints etc.).\\
		$I :=$ The proposed tool's programmatic interface.\\
		$\psi_{min} :=$ Minimum size of the input code.\\
		$\psi_{max} :=$ Maximum size of the input code.\\
		$\delta_{min} :=$ Fingerprint matching threshold.\\
		\ENSURE 
		$T_v :=$ Database table containing verification results.\\
		
		\STATE  $P_{input}^{unique} =$ Select N $\langle c, p\rangle$ tuples from PostsDB\\
		~~~~~~~~~~~~ $\mid \psi_{min} \leq \psi \leq \psi_{max}$ and $p$ is unique.
		\STATE  $P_{input}^{dup} =$ Select $N$ $\langle c', p', r\rangle$ tuples from PostsDB\\
		~~~~~~~~~~~~ $\mid  \psi_{min} \leq \psi \leq \psi_{max}$  and $p'$ has a duplicate $r$.\\
		/*
		\texttt{Here, $p, p', r$ are posts, and $c, c'$ are code parts
			belonging to $p, p'$ respectively. $\psi$ is length of code $c$ or $c'$.}
		*/
		\FOR {Every code part $c \in P_{input}^{unique}$}
		\STATE  $M = I.findMatchingPosts(c, \delta_{min})$
		\FOR {Each match $m \in M$}
		\STATE  Save match info from $\langle c, p, \delta_{actual}, \psi, m\rangle$ into $T_v$
		\ENDFOR
		\ENDFOR
		
		\FOR{Every code part $c' \in P_{input}^{dup}$}
		\STATE $M' = I.findMatchingPosts(c', \delta_{min})$
		\FOR {Each match $m' \in M'$}
		\STATE Save match info from $\langle c', p', r, \delta_{actual}, \psi, m'\rangle$ into $T_v$
		\ENDFOR
		\ENDFOR
		
	\end{algorithmic}
\end{algorithm}

\subsection{Key questions considered in empirical evaluation}
\label{sec:verif_setup}
Primarily, following variables affect the overall performance of our tool:
\begin{enumerate}[label=\roman*)]
	\item Number and \textit{quality} of posts available in StackOverflow data dump. The \textit{quality} of a post would be considered good if it contains sufficient amount of properly written source code, and also has reliable meta-data such as score, classification tags etc.
	\item Ability of the code matching algorithm to identify correct matches.
	\item Size of the input source code that is being reviewed.
	\item Correct setting of various parameters of the tool and its algorithms.
\end{enumerate}

Contents of the posts available in StackOverflow data dump is fixed (at a given time). Therefore, when testing for the effectiveness of our tool we mainly experiment with the last three variables, i.e., parameters of the code matching algorithm, the input source code and the settings of various parameters of the tool. As such, following are some of the key questions around which we designed our empirical evaluation experiments:
\begin{enumerate}[label=\roman*)]
	\item What level of precision can be expected for the tool's output? Is there any correlation of precision with other metrics such as level of fingerprint match etc.?
	\item How does the choice of fingerprint matching threshold, $\delta$, affect tool's ability to detect relevant posts?
	\item Does the size of input code size, $\psi$, affect quality of tool's output? In other words, does the input code size, $\psi$, affect the number of relevant posts and the defectiveness score reported by the tool?
	\item What is the minimum value of $\delta$ which is required to correctly detect a post that has a code part matching, up to a given \%age, with the input code?
\end{enumerate}

Two parameters play an important role when using the proposed tool: a) The fingerprint match threshold, $\delta$, and b) The size, $\psi$, of input source code. Answers to the above questions would help in suitably tuning the tool's parameters for getting the best results.

For carrying out empirical evaluation experiments we developed the necessary programs using Java programming language. Algorithm-\ref{alg:verification_setup} shows the steps performed by this program.

\subsection{Empirical evaluation}
\label{sec:exp_details}
Here we discuss the details of important experimental measurements that we performed to evaluate: \textbf{a)} how effective is the proposed tool in identifying StackOverflow posts which are relevant for reviewing input source code, and \textbf{b)} effectiveness of the tool when used on real projects by the programmers.

\subsubsection{Correlation between degree of code mismatch and $\delta_{actual}$}
An important question we wanted to answer is that for a given amount of textual difference, $\psi_{diff}$, between two samples $C_1$ and $C_2$ of source code, what is the actual observed amount of fingerprint match $\delta_{actual}$. An answer to this question would offer a guideline for selecting the fingerprint match threshold for different code matching scenarios.

For determining the answer we measured the effect of $\psi_{diff}$ on $\delta_{actual}$. The textual difference $\psi_{diff}$ is measured in terms of aggregate length of the text regions that do not match in the two source code samples.

The code samples $C_1$ and $C_2$ are strings synthesized from an alphabet $\Theta_{code}$. For introducing mismatching regions into one of the samples we used short strings synthesized from another alphabet $\Theta_{noise}$. The locations of mismatching regions in a code sample were chosen randomly.

\begin{figure}
	\centering
	\includegraphics[width=0.5\linewidth]{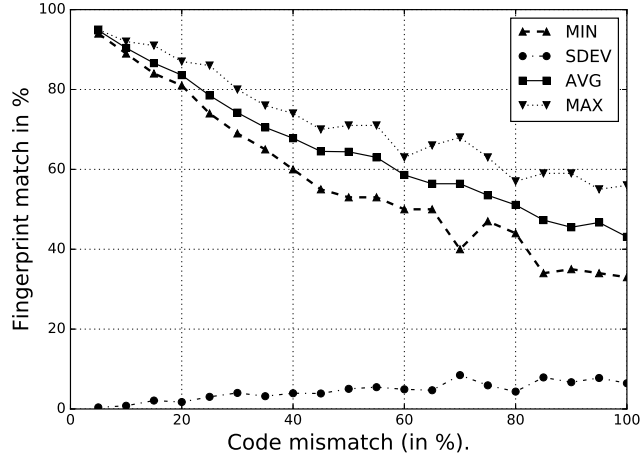}
	\caption{Fingerprint match vs. Degree of code mismatch.}
	\label{fig:fperror_by_altpct}
\end{figure}

Fig. \ref{fig:fperror_by_altpct} depicts the observed correlation between $\delta_{actual}$ and the degree of code mismatch $\psi_{diff}$. An important observation is that:

\begin{tcolorbox}
On average, it is possible to have $\approx 40\%$ match of fingerprints for two completely different code samples. For detecting two code samples which are at least $50\%$ similar, their fingerprints match should be $\geq 60\%$.
\end{tcolorbox}

\subsubsection{Code matching performance}
\label{sec:code_matching_accu}
An important part of our tool is the algorithm that we employ for handling contextual search of source code parts in \texttt{PostsDB}. Effectiveness of the proposed tool depends to a great extent on accuracy of the method used for source code matching. This is because the relevance of search results is proportional to the degree to which an input source code matches with code parts\footnote{A post has one or more pieces of code embedded in it. Each such embedded piece of code is called a ``part''. A ``part'' may or may not be a complete program.} found in \texttt{PostsDB}. For performing source code matching we have employed a well known technique called Winnowing \cite{schleimer2003winnowing} which is a local algorithm for ``document'' fingerprinting. The code matching effectiveness and limitations that such algorithm has will also be applicable to our tool. 

In order to evaluate and compare the performance of proposed tool for various scenarios in our experiments we use the following metrics:
\begin{enumerate}[label=\roman*)]
	\item We define a metric, $\Omega$, that we call matching ratio, as in Equation-\ref{eq:matching_ratio}. 
	\item We use the standard metric of \textit{precision}, as defined in Equation-\ref{eq:precision}.
\end{enumerate}

\begin{equation}
\label{eq:matching_ratio}
\Omega = \frac{\sum_{i=1}^{N} |M_i|}{N}
\end{equation}

\begin{equation}
\label{eq:precision}
\xi = \frac{tp}{tp+fp}
\end{equation}

Here, $|M_i|$ is the number of matched posts for $i^{th}$ input, and $N$ is the total number of inputs. $M$ (or $M'$) and $N$ are as described in Algorithm-\ref{alg:verification_setup}. $\Omega$ gives an average measure of the matches found per input. $tp$ is the number of true positive results, and $fp$ is the number of false positives observed in an experiment. A string token matching algorithm was used to identify \textit{false positive} code matches (we also manually verified several randomly selected false positives).

To ensure the correct implementation of Winnowing based code matching algorithm we have evaluated its performance in two scenarios:
\begin{enumerate}[label=\roman*)]
	\item When code input to the tool was substantially similar to the code present in a post.
	\item When code input to the tool matched only partially with the code present in a post.
\end{enumerate}

In our experiments we observed that for both of the above mentioned scenarios the expected matches were always identified by the tool. That is, \textit{false negatives} were zero. Section-\ref{sec:key_obs} gives a detailed discussion of the results.

\begin{figure*}[!t]
	\centering
	\begin{subfigure}[t]{0.3\textwidth}
		\centering
		\includegraphics[height=1.5in]{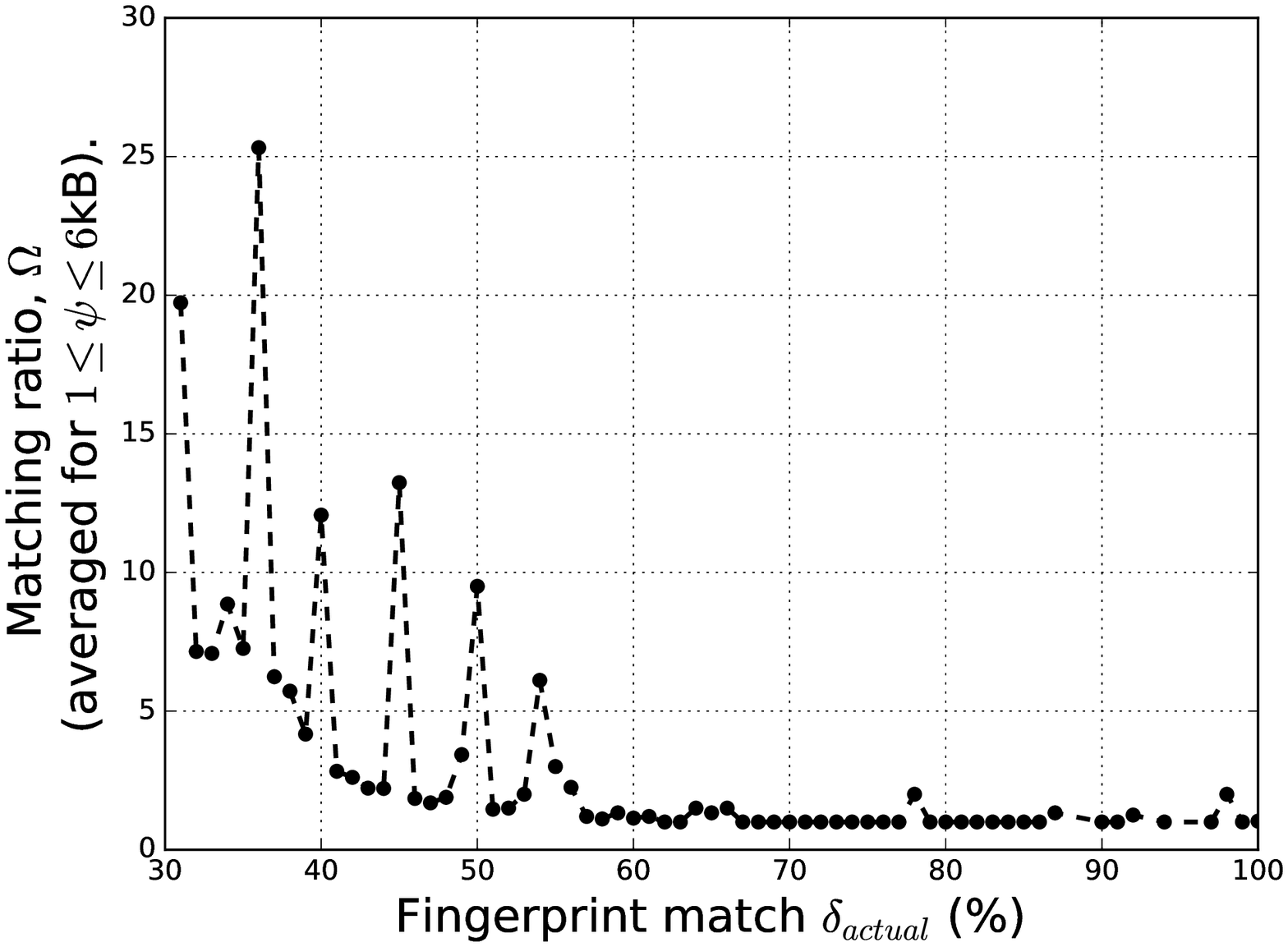}
		\caption{Effect of fingerprint match, $\delta_{actual}$.\label{fig:omega_vs_delta}}
	\end{subfigure}
	~
	\begin{subfigure}[t]{0.3\textwidth}
		\centering
		\includegraphics[height=1.5in]{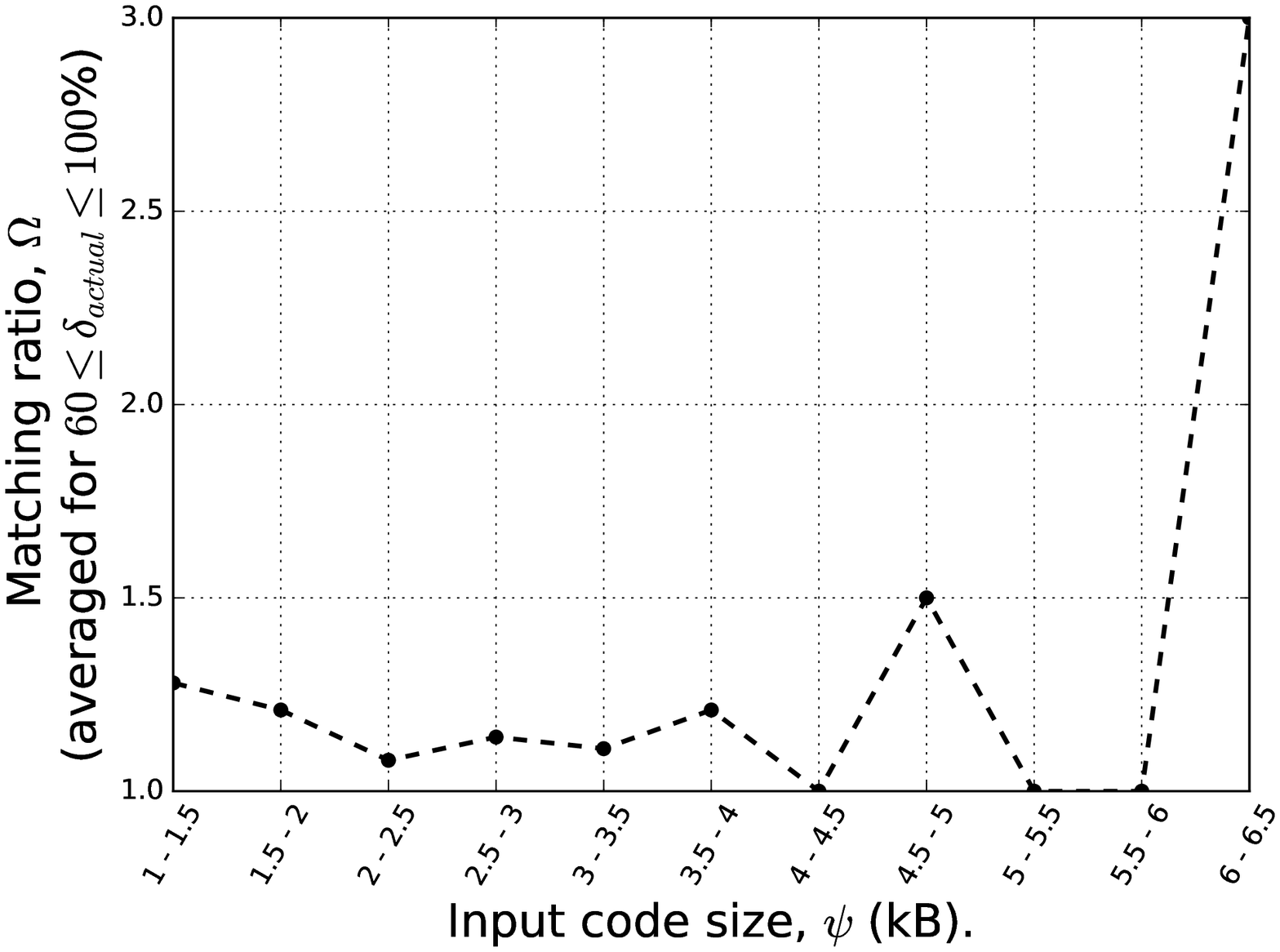}
		\caption{Effect of input code size, $\psi$. \label{fig:omega_vs_psi}}
	\end{subfigure}
	~
	\begin{subfigure}[t]{0.3\textwidth}
		\centering
		\includegraphics[height=1.5in]{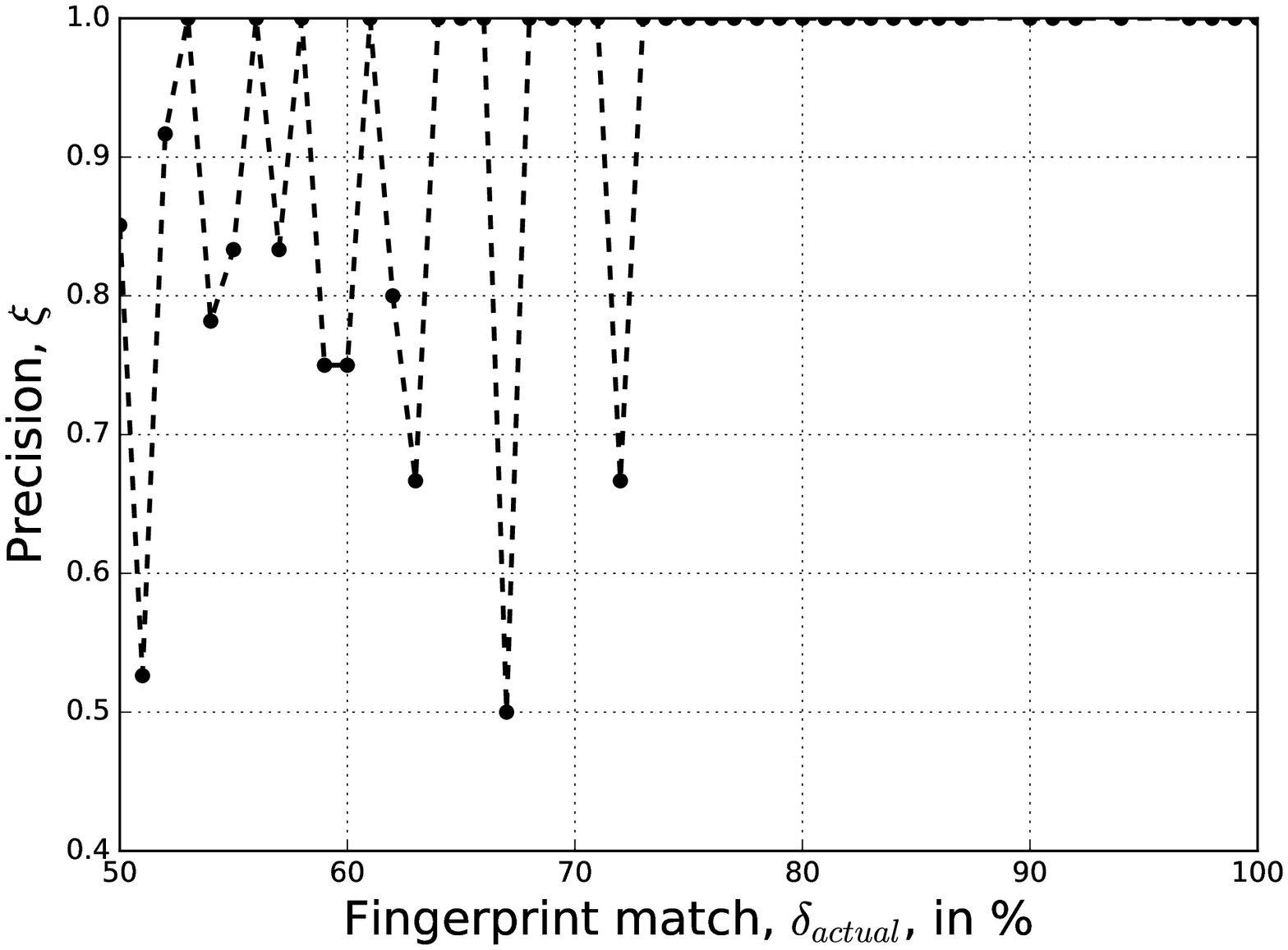}
		\caption{Precision, $\xi$. \label{fig:precision}}
	\end{subfigure}
	
	\caption{Precision and matching ratio performance (when input taken from unique posts).}
	\label{fig:omega_delta_psi}
\end{figure*}

\subsubsection{Evaluation using input code taken from StackOverflow posts}
\label{sec:verif_dup_posts}

Rationale for this approach is based on a pattern that we observed in posts on StackOverflow.
We found that there are sufficient number\footnote{We composed a query (available here: \url{https://data.stackexchange.com/stackoverflow/query/620299}) to estimate the number of posts on StackOverflow which have a tag \texttt{Java} and have been marked to have a possible duplicate. The query reports about 398409 posts to be duplicate as of Nov 2017.}
of posts on StackOverflow which are flagged by the moderators (at StackOverflow) as duplicate of some other post. A rigorous crowd-sourced review approach\footnote{Please refer to this document: \url{http://meta.stackexchange.com/questions/10841/how-should-duplicate-questions-be-handled} and the database schema here: \url{https://data.stackexchange.com/stackoverflow/query/new}} is employed by StackOverflow before labeling a post as duplicate. Availability of such programmer verified $\langle original, duplicate\rangle$ pairs of posts in StackOverflow data offers us very good quality test data set for evaluation of our tool. Correctness of our tool is verified if the tool reports in its results both the original post and its duplicate(s), for an input which included the code from only the original post.

Following are the steps that we performed in this scenario:

\begin{enumerate}[label=\roman*)]
	\item Select a random set  $P_{test}$ of $\langle p^{orig}, p^{dup}\rangle$ tuples of posts from \texttt{PostsDB}.
	\item For each tuple $\langle p^{orig}_i, p^{dup}_i\rangle \in P_{test}$ extract the code parts $C_i$ from $p^{orig}_i$.
	\item Run the tool on $C_i$ to get a set $M$ of relevant posts.
	\item If the set $M$ contains both $p^{orig}_i$ and $ p^{dup}_i$, then we deduce that the tool worked correctly to identify the desired posts.\\
\end{enumerate}

Results are depicted in Fig. \ref{fig:omega_delta_psi} and \ref{fig:omega_delta_psi_dup} and further discussed in Section-\ref{sec:key_obs}.

\subsubsection{Evaluation by programmers on real projects}
\label{sec:verif_manual}
The tool was also evaluated by a group of programmers who were actively involved in developing code for real projects. Evaluation was performed on source code of open source projects available on GitHub. Table-\ref{tab:real_proj_eval} shows the relevant information about those projects. Following are the steps in our evaluation process:

\begin{enumerate}[label=\roman*)]
	\item Select a set, $G$, of open source projects from GitHub repositories. \textit{In our current evaluation set up we considered mostly Java source code.}
	
	\item For each project $g \in G$, randomly select a set $S_{g}$ of source files.
	
	\item Run the tool for each source file $s \in S_{g}$ to produce a set $M$ of relevant posts.
	
	\item A team of experienced programmers checks the relevance of posts in $M$.
	
\end{enumerate}

We developed a program to fetch source code files from open source GitHub repositories. The program made use of GitHub search API to search and fetch the source files from repositories. We downloaded only those source files whose size was between 1 to 10kB. Other than source code file size we did not impose any specific constraints on size of teams or the size of code bases for respective projects.

Observations from this evaluation are discussed in Section-\ref{sec:github_code_verification}.

\begin{table}[ht]
	\centering
	\caption{Evaluation of the tool on GitHib OSS projects.}
	\label{tab:real_proj_eval}
	\begin{tabular}{c|c|c|c}
		\hline
		\textbf{No. of repositories} & \textbf{Average repository size in MB} & \textbf{Average No. of contributors} & \textbf{Main language} \\ \hline
		303 & 4.34 & 2.47 & Java \\ \hline 
		21 & 3.24 & 1.81 & JavaScript \\ \hline 
		6 & 6.10 & 0.67 & HTML \\ \hline 
		5 & 18.19 & 7.60 & C++ \\ \hline 
		4 & 2.14 & 1.00 & CSS \\ \hline 
		3 & 16.30 & 1.33 & PHP \\ \hline 
		3 & 89.84 & 1.00 & C \\ \hline 
		
	\end{tabular}
\end{table}

\subsubsection{Defectiveness score validation}
\label{sec:verif_sentiment}
As described in Section-\ref{sec:deriving_defectiveness}, determining defectiveness score for the input code is relatively simple once the correct matching posts have been identified. To assess relevance of defectiveness scores as determined by our tool we have compared the defectiveness score $\alpha^p$ of a post $p$ determined by our tool with the score $\alpha^p_{reference}$ calculated for the same post using different NLP based tools such as CoreNLP\cite{corenlp2014} and VADER\cite{hutto2014vader}. We adopted the following approach for carrying out this comparison:\\
For each matched post $p$ we do:
\begin{enumerate}[label=\roman*)]
	\item Calculate defectiveness score $\alpha^p$ as described in Algorithm-\ref{alg:main_algo} and \ref{alg:defectiveness}.
	\item Calculate\footnote{Calculated using the implementation of CoreNLP from: https://stanfordnlp.github.io/CoreNLP/ and VADER from https://github.com/cjhutto/vaderSentiment} defectiveness score $\alpha^p_{reference}$ for the \underline{narrative text} found in post $p$.
	\item Compare the values $\alpha^p_{reference}$ with $\alpha^p$ to find the overall degree of concurrence. We calculate the fraction $\Gamma$ as:
	\begin{equation}
	\label{eq:sentiment_match}
	\Gamma = \frac{\text{No. of posts for which } \alpha^p \text{ matches } \alpha^p_{reference}}{\text{Total number of posts.}}	
	\end{equation}		
\end{enumerate}

\section{Key observations from evaluation experiments}
\label{sec:key_obs}
A major goal of our tool is to determine defectiveness score for code parts present in StackOverflow posts so that we can estimate the defectiveness score for a similar input source code. Also, identifying StackOverflow posts which contain source code snippets that sufficiently resemble the input code remains an important step in our approach. As such we designed our experiments to evaluate our tool for code matching performance as well as correctness of reported defectiveness scores for the input code. Evaluation has been carried out for a wide range of inputs under different parameter settings of the tool as discussed next.

\begin{table*}[ht]
	\centering
	\caption{Key statistics for evaluation using input from StackOverflow posts.}
	\label{tab:so_verification_stats}
	\begin{tabular}{l|c|c}
		\hline
		& \multicolumn{2}{c}{\textbf{VALUE} (for $\delta_{actual} \geq 50\%$)} \\ \cline{2-3}
		\textbf{ITEM} & \textbf{For unique inputs} & \textbf{For inputs with a duplicate} \\ 
		\hline
		Total number of input files &	500	& 500\\ \hline
		Input files having a match &	500 & 500	\\ \cline{2-3}
		&  \multicolumn{2}{c}{\textbf{$\langle$Average, Minimum, Maximum, Std. Deviation$\rangle$}} \\ \hline
		Fingerprint match, $\delta_{actual}$ (in \%) & $\langle$83.71, 50, 100, 21.62$\rangle$ & $\langle$64.26, 50, 100, 17.34$\rangle$ \\ \hline
		Input code size, $\psi$ (kB) & $\langle$2.0, 1.0, 6.2, 1.0$\rangle$ & $\langle$2.0, 1.0, 6.4, 0.9$\rangle$ \\ \hline
		Matching ratio, $\Omega$ & $\langle$1.72, 1.00, 141.00, 6.49$\rangle$ & $\langle$7.61, 1.00, 758.00, 43.73$\rangle$	\\
		\hline
		Precision, $\xi$ (in \%)&	$\langle$96.78, 66.67, 100, 8.48$\rangle$ & $\langle$97.77, 50.55, 100, 7.98$\rangle$ \\
		\hline
	\end{tabular}
\end{table*}

\subsection{Observations from empirical evaluations}
\label{sec:obs_emp_eval}

\subsubsection{Effect of $\delta$ and $\psi$ on matching ratio $\Omega$}
We experimented with two scenarios: \textbf{i)} where input code was chosen from existing posts that did not have a duplicate post, and \textbf{ii)} where input code was chosen from posts that had duplicate posts. Important statistics seen from our experiments are summarized in Table-\ref{tab:so_verification_stats}.

Fig. \ref{fig:omega_delta_psi} depicts observations for scenario \textbf{i} where we see following salient points:
\begin{itemize}
	\item The cut-off point for fingerprint match $\delta_{actual}$ after which the matching ratio $\Omega$ remains stable at its expected value of 1 is $\approx 57\%$. The average value of precision is about 96\% for $\delta_{actual}\geq 60\%$. In other words, number of false positives becomes very small for $\delta_{actual} > 60\%$.
	\item $\Omega$ and number of false positives rise sharply as the value of $\delta_{actual}$ falls below 60\%.
	\item Overall average precision, $\xi$, of the results remain at $\approx 93\%$ (trend is shown in Fig. \ref{fig:precision}).
\end{itemize}

\begin{figure*}
	\centering
	\begin{subfigure}[b]{0.3\textwidth}
		\includegraphics[height=1.5in]{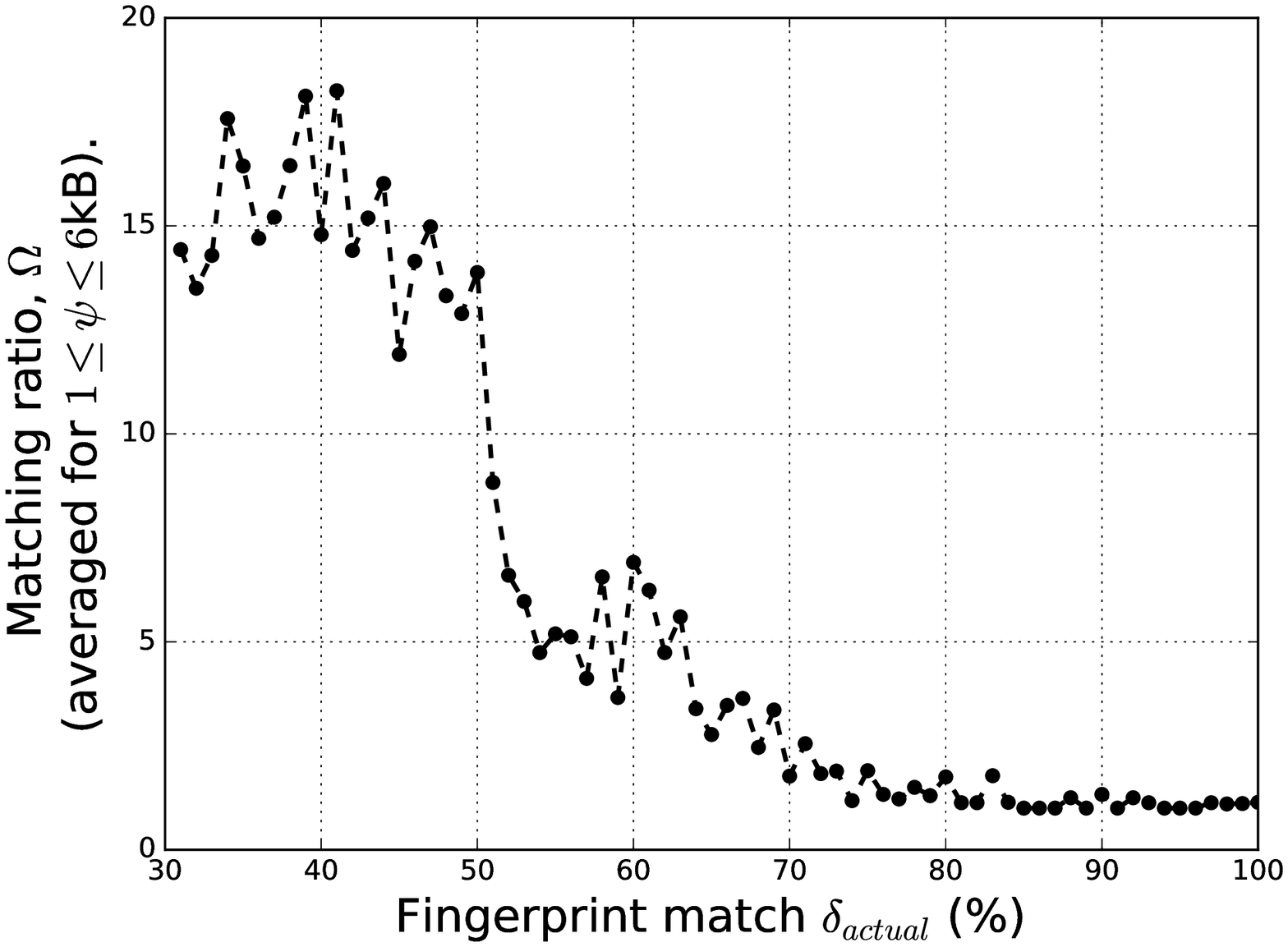}
		\caption{Effect of fingerprint match, $\delta_{actual}$.\label{fig:omega_vs_delta_dup}}
	\end{subfigure}    
	~
	\begin{subfigure}[b]{0.3\textwidth}
		\includegraphics[height=1.5in]{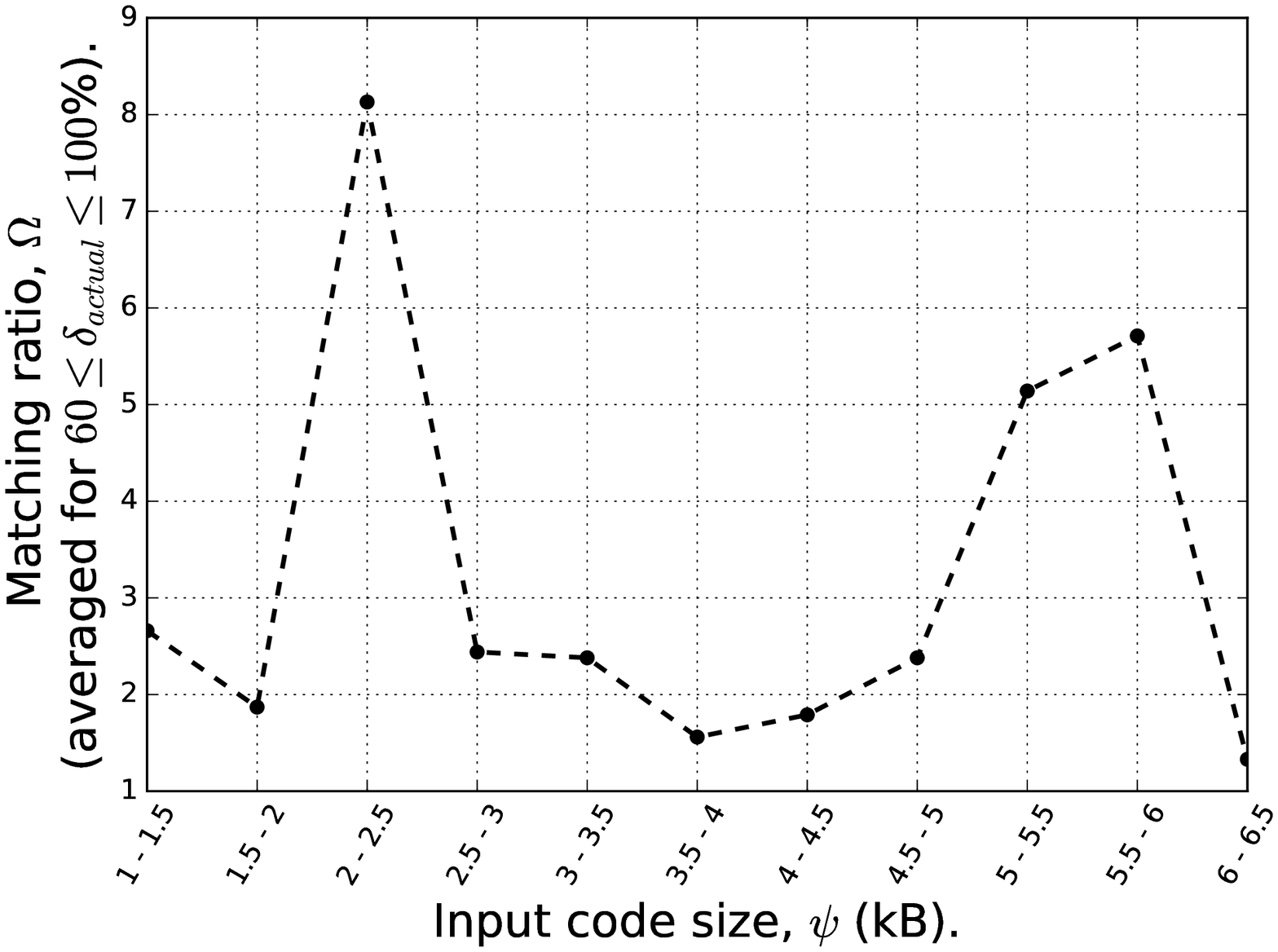}
		\caption{Effect of input code size, $\psi$. \label{fig:omega_vs_psi_dup}}
	\end{subfigure}   
	~
	\begin{subfigure}[b]{0.3\textwidth}	%
		\includegraphics[height=1.5in]{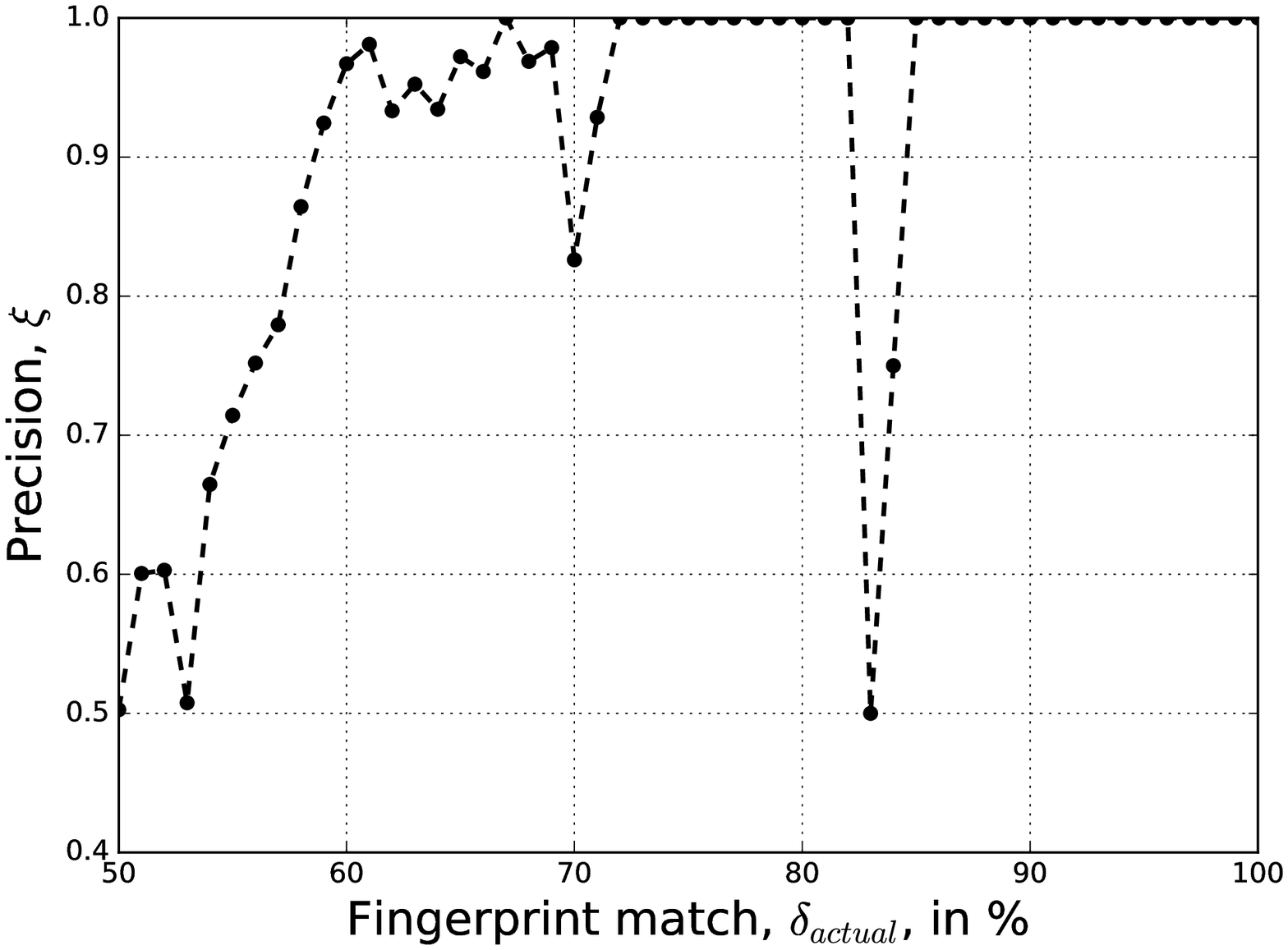}
		\caption{Precision, $\xi$.\label{fig:precision_dup}}
	\end{subfigure}   
	\caption{Precision and matching ratio performance (when input taken from posts having a duplicate).}
	\label{fig:omega_delta_psi_dup}
\end{figure*}

Fig. \ref{fig:omega_delta_psi_dup} shows observations for scenario \textbf{ii}, and the salient points are:
\begin{itemize}
	\item $\Omega$ settles at about 2 (expected value) when fingerprint match $\delta_{actual} \geq 72\%$. The average precision also remain about 97\% for $\delta_{actual} \geq 70\%$.
	\item Overall average precision, $\xi$, in this case is $\approx 91\%$ (trend is shown in Fig. \ref{fig:precision_dup}).
\end{itemize}

It is difficult to deduce any clear correlation between the size, $\psi$, of input code and the matching ratio, $\Omega$, as seen from Fig. \ref{fig:omega_vs_psi} and \ref{fig:omega_vs_psi_dup}. \\
\textbf{Possible reasons for observed trend:} A near independence of $\Omega$ from size $\psi$ of input code, is likely due to the large standard deviation observed in the sizes of source code present in StackOverflow posts. As we see in Table-\ref{tab:posts_stats}, the value of standard deviation is as large as the mean. Such a large value of standard deviation implies that there is significant dispersion in code sizes. Therefore, the variation of input code size seems to have little effect on the matching ratio.

\subsubsection{Evaluation on real projects}
\label{sec:github_code_verification}
Effectiveness of the proposed tool was also tested by using input source code from open source project repositories at GitHub. We used 370 source code files (size between 1 - 10kB each) from 345 different repositories hosted at GitHub. Key statistics from our testing are summarized in Table-\ref{tab:github_stats}.

\begin{table*}[ht]
	\centering
	\caption{Key statistics for GitHub code evaluation.}
	\label{tab:github_stats}
	\begin{tabular}{l|c|c}
		\hline
		& \multicolumn{2}{c}{\textbf{VALUE}} \\ \cline{2-3}
		\textbf{ITEM} & \textbf{For $\delta_{actual} \geq 30\%$} & \textbf{For $\delta_{actual} \geq 50\%$} \\ 
		\hline
		
		Total number of input files &	370	& 370 \\ \hline
		Input files having a match &	50 & 2	\\ \hline
		&\multicolumn{2}{c}{\textbf{$\langle$Average, Minimum, Maximum, Std. Deviation$\rangle$} }\\ \hline
		Fingerprint match, $\delta_{actual}$ (in \%)	&	$\langle$34.2, 31.0, 53.0, 3.6$\rangle$	 & $\langle$51.4, 50.0, 53.0, 0.97$\rangle$ \\ \hline
		Input code size, $\psi$ (kB) &	$\langle$7.3, 1.1, 9.9, 2.3$\rangle$ & $\langle$9.6, 9.2, 9.9, 0.35 $\rangle$ \\ \hline
		Matching ratio, $\Omega$&	$\langle$20.6, 1.0, 208.0, 38.2$\rangle$ & $\langle$5.5, 5.0, 6.0, 0.5$\rangle$ \\
		\hline
		Precision, $\xi$ (in \%)&	$\langle$97.34, 89.66, 100, 3.68$\rangle$ & $\langle$100, 100, 100, 0.0$\rangle$ \\
		\hline
	\end{tabular}
\end{table*}

\begin{figure*}
	\centering
	\begin{subfigure}[t]{0.3\textwidth}
		\includegraphics[height=1.5in]{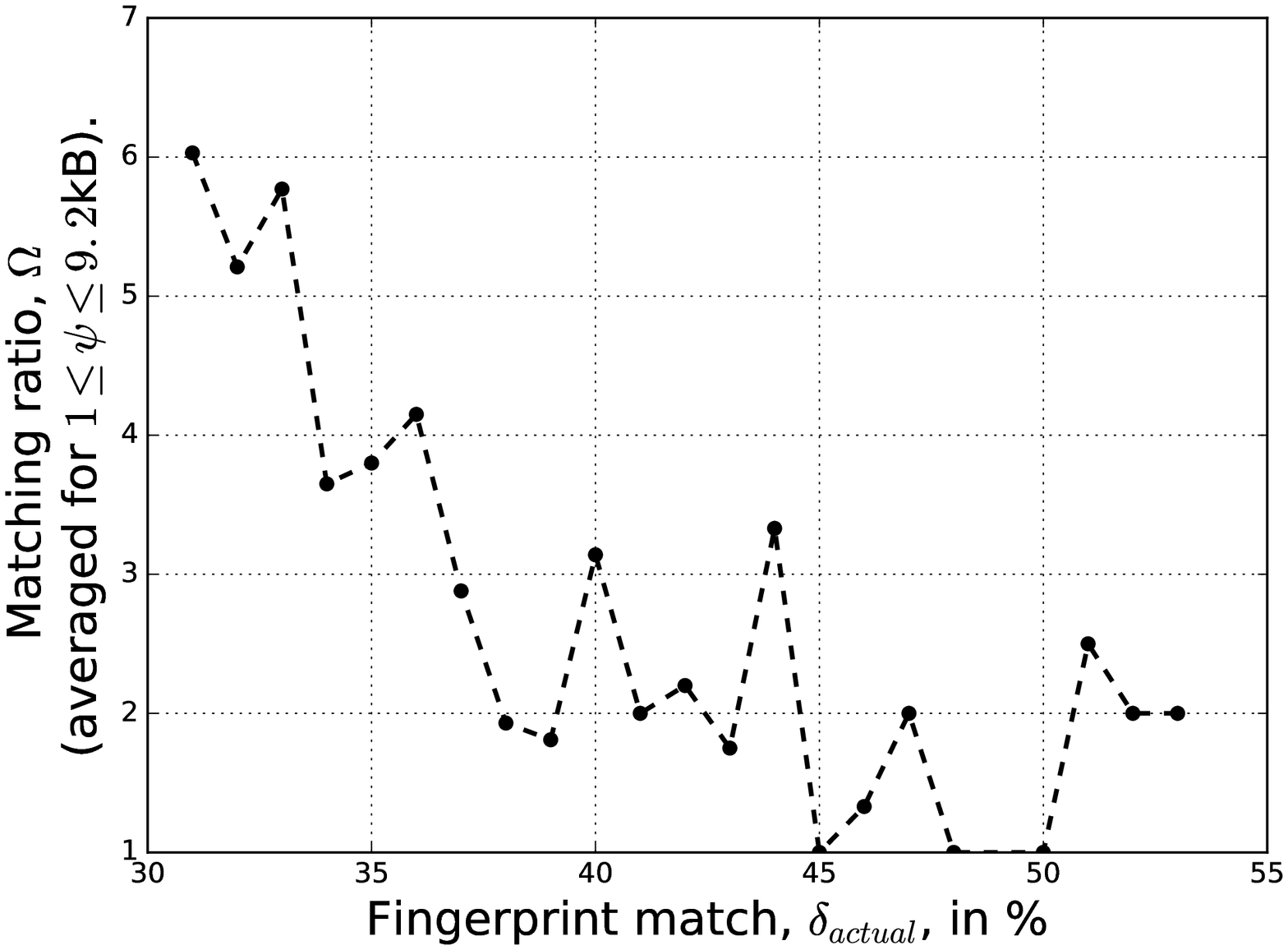}
		\caption{Effect of fingerprint match, $\delta_{actual}$ ($\geq 30\%$ only.) \label{fig:actual_delta_vs_omega}}
	\end{subfigure}
	~
	\begin{subfigure}[t]{0.3\textwidth}
		\includegraphics[height=1.5in]{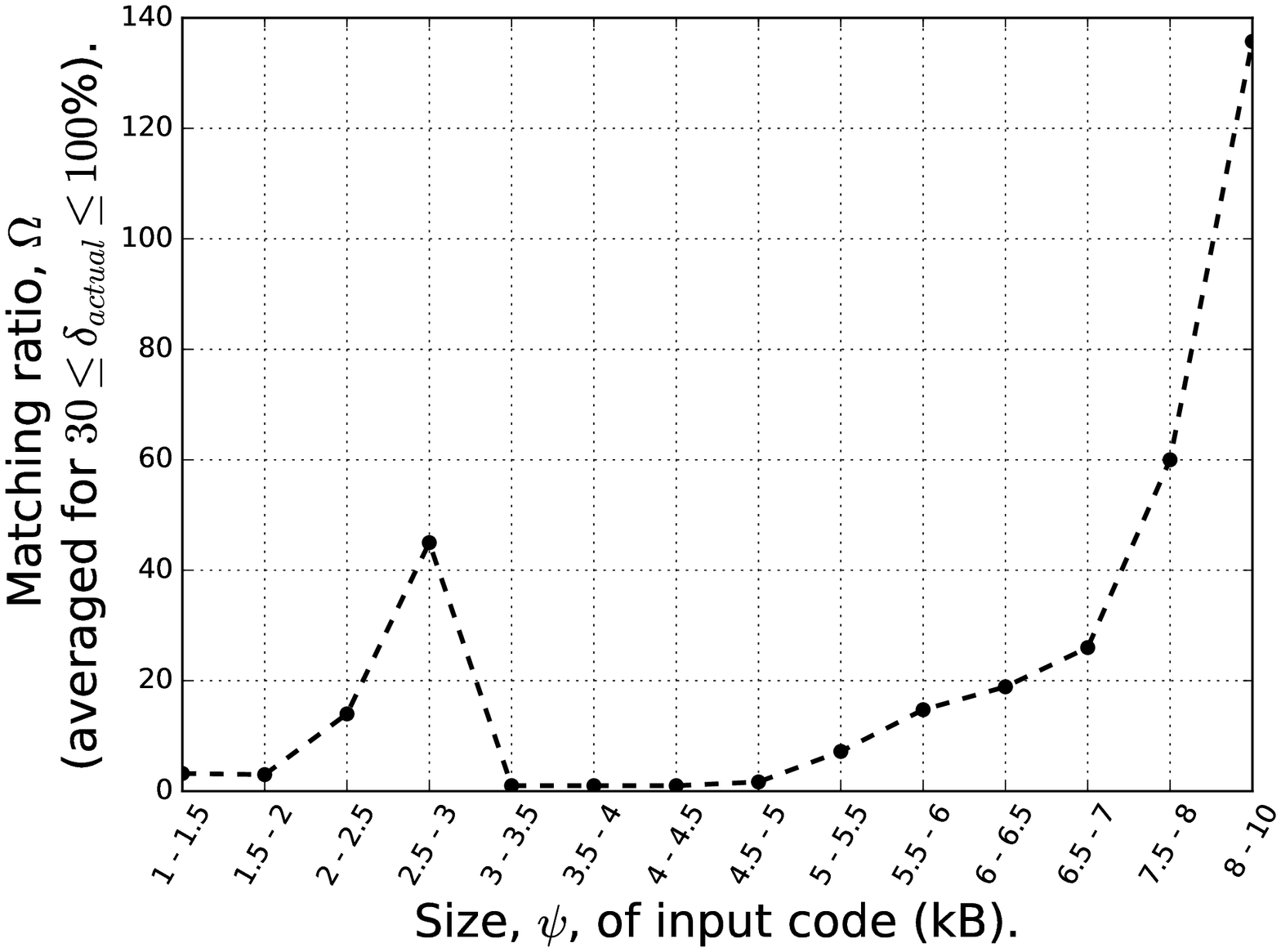}
		\caption{Effect of input code size, $\psi$. \label{fig:actual_delta_vs_psi}}
	\end{subfigure}
	~
	\begin{subfigure}[t]{0.3\textwidth}
		\includegraphics[height=1.5in]{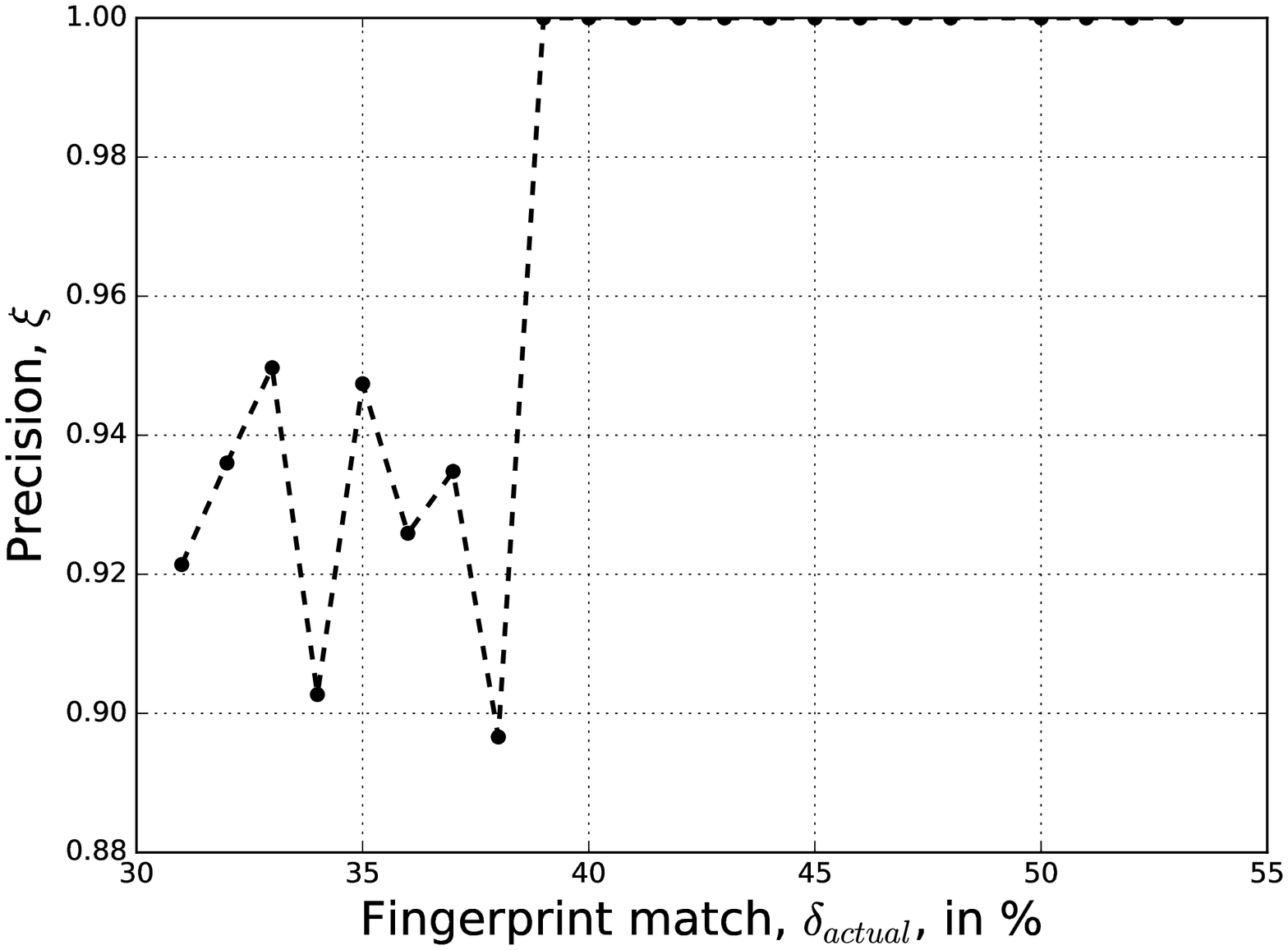}
		\caption{Precision, $\xi$.\label{fig:git_precision_dup}}
	\end{subfigure}
	\caption{Evaluation with input code taken from GitHub repositories.\label{fig:manual_verification}}
\end{figure*}

Fig. \ref{fig:manual_verification} depicts the variation in matching ratio, $\Omega$, w.r.t. observed fingerprint match, $\delta_{actual}$, and size, $\psi$, of the input source code. The observed general trends are similar to the ones seen in previous scenarios, though the absolute values are different. Notable observations are: 
\begin{itemize}
	\item The largest value observed for fingerprint match, $\delta_{actual}$, is about 53\%.
	\item Less than 1\% of the input files find a relevant matching post having $\delta_{actual} \geq 50\%$.
	\item For $\delta_{actual} \geq 50\%$ the value of $\Omega \leq 3$.
	\item Average precision, $\xi$, of the results is about 97\%.
\end{itemize}

Interestingly, our observation that very few input files find a sufficiently relevant/matching post seems to validate the results reported by Bogdan et. al. in \cite{stackoverflow-github} where one of their observation is:
\begin{quote}
	``Active GitHub committers ask fewer questions on StackOverflow than others.''
\end{quote}

\subsubsection{Correctness of defectiveness score}
\label{sec:obsev_sentiment}
A post's \texttt{SCORE} value (available in the post's mete-data) indicates its acceptance by the professional programmers. Likelihood of a post being representative of a real and genuine programming issue/scenario is proportional to the \texttt{SCORE} value achieved by the post. This is the main reason we have used \texttt{SCORE} value of a post for calculating its defectiveness score in Algorithm-\ref{alg:defectiveness}. Table-\ref{tab:defect_score} shows important statistics about the \texttt{SCORE} values observed for different types of posts available in StackOverflow data dump that we used.

\begin{table}[ht]
	\centering
	\caption{Statistics about calculated defectiveness scores.}
	\label{tab:defect_score}
	\begin{tabular}{l|c|c|c}
		
		\hline
		& \multicolumn{3}{c}{\textbf{ Item value for different post types}} \\ \cline{2-4}
		\textbf{ Item} & \textbf{ Questions} & \textbf{ Replies} & \textbf{ Accepted replies} \\ \hline		
		
		Average \texttt{SCORE} value & 4.23 & 4.08 &  7.11\\ \hline 
		Minimum \texttt{SCORE} value & -147 & -58 &  -54\\ \hline 
		Maximum \texttt{SCORE} value & 9432 &  11055 &  11055\\ \hline 
		Std. Deviation of \texttt{SCORE} & 28.49 &  26.8 &  41.08\\ \hline 
		$\Gamma$ in \% from Eq.-(\ref{eq:sentiment_match}) &  42 &  29 &  32\\ \hline
		
	\end{tabular}%
\end{table}

We observe that $\Gamma \approx 42\%$ (Equation-\ref{eq:sentiment_match}) for the posts of type \textit{question}. That is, the defectiveness scores $\alpha$ and $\alpha_{reference}$ matched for about $42\%$ of such posts. This number was 29\% and 32\% for all posts of type \textit{reply} and \textit{accepted reply} respectively. These numbers were observed when we set the $\alpha_{threshold}$ value in Algorithm-\ref{alg:defectiveness} equal to average value of \texttt{SCORE} observed in the StackOverflow data.

\textbf{Possible reasons for observed trend:} We analysed the low values for $\Gamma$ by manually examining a subset of recommendations produced by our tool. We observed that the text narrative available in StackOverflow posts makes use of vocabulary which is specific to the domain of software development. A polar word/phrase in such a narrative and vocabulary may or may not be considered polar from normal English language perspective. For example consider the following sentence:

\begin{quote}
	\texttt{``The following piece of code takes a huge amount of memory and CPU, and takes very long to produce results.''}
\end{quote}

A narrative similar to the above sentence about a piece of code is highly likely to be considered negative. However, most existing NLP based tools would label this sentence as either a ``neutral'' or ``positive'' comment about the code it refers to. For instance, as of December 2017, the CoreNLP\cite{corenlp2014} Sentiment Analysis tool (available live at \url{http://corenlp.run}) labels this sentence as positive. Similarly, Vader\cite{hutto2014vader} too labels it as positive.

We manually checked a random subset of instances where the defectiveness scores $\alpha$ and $\alpha_{reference}$ did not match. In all such cases the defectiveness score assigned by our algorithm was found to be correct.

\subsection{Threats to validity}
\label{sec:threats}
The basic premise on which our tool works is that: 
\begin{itemize}
	\item the code present in a question post or a poorly rated reply post is highly likely to be of poor quality, and
	\item  the code present in an accepted answer post or a highly rated reply post is highly likely to be of good quality.
\end{itemize}

Often there are more than one correct ways to write code for certain scenarios. Sometimes a solution coded in a particular style gets higher rating/acceptance\cite{stackoverflow-repute,stackoverflow-myths} mainly because it matches the personal/organizational coding style of the programmer who posted the original question. In such cases even though the code present in low rated reply posts may be defect free, there are chances that such code may get a poor defectiveness score in relative terms.

Another possible shortcoming may be in the manner by which the \texttt{Code Matching Query Handler} processes input code. Irrespective of whether the tool uses an input code file as a single code block or breaks it into smaller basic blocks, there remains a possibility of leaving out relevant matches available in \texttt{PostsDB}. The reason why this may happen is as follows: Code present in a post may be anything from a complete and compilable unit (e.g. a Java class definition) pasted as-is, to a small fragment of it (e.g. a while loop from a function definition). Thus, if the tool uses a full compilation unit as-is to search for matches in \texttt{PostsDB}, then those posts which contain only a small fragment from such code may not show up as a match. Similar situation occurs for the reverse scenario, i.e., when input is a small fragment but a post contains a complete program.

Also, in our current implementation we do not pre-process the source code to neutralize the effect of coding style variations such as different naming style for variables etc. Incorporating such pre-processing can enhance the effectiveness of our tool.

Lastly, limitations and parameter settings of the code fingerprinting algorithm\cite{schleimer2003winnowing} that we use can also affect the quality of results of our tool. For example, when implementing this fingerprinting algorithm the choice of a hash function that the algorithm uses affects the quality of code matching.

\section{Conclusion}
Lowering the cost of creating and maintaining good quality software is undoubtedly an important goal in software development. Researchers have shown that improving code quality can significantly lower the overall cost of software development and maintenance. In this paper we have proposed a tool that helps a programmer in writing better code by pointing out potentially buggy portions of the code. 

The discussions available on professional programmer support forums such as StackOverflow are used by our tool for assessing defectiveness of input code. Defectiveness of the input source code is estimated by inferring the ``sentiment'' of discussion posts that contain any code that is similar to the input code. A key idea in our approach is the use of a document fingerprinting technique for efficient and accurate comparison of source code when searching for similar code in discussion posts. The document fingerprinting techniques have been very successfully used in source code plagiarism detection tools. They make a good choice for us because the code matching scenario that our tool faces is similar to the source code plagiarism detection.

Efficacy of the proposed tool has been verified by checking the correctness of results along three important dimensions: a)
by measuring code matching accuracy, b) by verifying results for known $\langle input, output\rangle$ pairs, and c) by verifying the results for code taken from real projects such as from GitHub.

Our experiments have shown that the document fingerprinting based search approach performs well in identifying relevant posts that contain code similar to an input code. We have shown that our tool performs better than existing NLP based techniques for determining defectiveness of a given piece of source code.

\bibliographystyle{IEEEtran}
\bibliography{main}

\end{document}